\documentclass[conference]{IEEEtran}
\IEEEoverridecommandlockouts
\usepackage{cite}
\usepackage{amsmath,amssymb,amsfonts}
\usepackage{algorithmic}
\usepackage{graphicx}
\usepackage{textcomp}
\usepackage{xcolor}
\usepackage{hyperref}
\usepackage{booktabs}
\usepackage{threeparttable}
\def\BibTeX{{\rm B\kern-.05em{\sc i\kern-.025em b}\kern-.08em
    T\kern-.1667em\lower.7ex\hbox{E}\kern-.125emX}}

\begin{document}

\title{EchoMark: Acoustic Environment Matching with Watermarked Room Impulse Response}

\author{
\IEEEauthorblockN{
Chenpei Huang\IEEEauthorrefmark{1},
Lingfeng Yao\IEEEauthorrefmark{1} \thanks{C. Huang and L. Yao share first authorship. M. Pan is the corresponding author. The work of C. Huang, L. Yao and M. Pan were supported in part by the US National Science Foundation under grants CNS-2107057, CNS-2318664, CSR-2403249, and CNS-2431596. The work of K. Lee was supported by National Research Foundation of Korea (NRF) grant funded by the Korea government(MSIT) (No. RS-2024-00463802).},
Lening Wang\IEEEauthorrefmark{2},
Lan Zhang\IEEEauthorrefmark{3},
Xun Chen\IEEEauthorrefmark{4},
Kyu In Lee\IEEEauthorrefmark{1},
and Miao Pan\IEEEauthorrefmark{1}
}

\IEEEauthorblockA{\IEEEauthorrefmark{1}University of Houston,
\IEEEauthorrefmark{2}Prairie View A\&M University,
\IEEEauthorrefmark{3}Clemson University,
\IEEEauthorrefmark{4}Independent Researcher}

\IEEEauthorblockA{
\{chuang30, lyao24, klee48, mpan2\}@uh.edu,
lenwang@pvamu.edu, lan7@clemson.edu, xunchen@outlook.com
}
}

\maketitle

\begin{abstract}
Recent advances in acoustic environment matching (AEM) allow creators to edit room acoustic effects using only a reference reverberant recording, enabling applications such as voice dubbing and immersive auditory virtual reality. However, such easy ``relocation” of audio environments raises security concerns, including realistic voice spoofing. We propose EchoMark, the first watermarked AEM framework that proactively protects and augments room impulse responses (RIRs). EchoMark adopts an EWG-D architecture with physics-inspired inductive biases to exploit watermark capacity that is imperceptible to human listeners. The embedded watermark remains detectable after convolution with arbitrary speech. Experiments show that EchoMark achieves competitive AEM quality among RIR reconstruction models while uniquely providing near-perfect watermark detection performance.
\end{abstract}

\begin{IEEEkeywords}
watermark, impulse response, acoustic environment matching
\end{IEEEkeywords}

\section{Introduction}
\label{sec:intro}
Audio recordings are a natural medium for conveying information and supporting interaction between humans and machines. When no additional context is specified, understanding an audio clip often focuses on its primary content, such as the spoken speech, sometimes even reduced to its transcript. However, rich contextual information is also embedded in the waveform, including the speaker’s identity, emotion, and surrounding environment.
Focusing on the acoustic environment, it is well known that room effects are determined by physical properties such as room shape, size, and surface materials \cite{aem-perceptual}. These factors jointly create perceptual cues that allow humans to recognize different environments. For example, a recording made in a bathroom sounds noticeably different from one recorded in an office, a distinction that can be identified by most non-expert listeners \cite{aem-perceptual}. Because room effects are closely tied to realistic and immersive listening experiences, transferring such effects to a new audio source has become an important creative technique, commonly referred to as acoustic environment matching (AEM).

One of the most direct examples of AEM is voice dubbing in the film industry, where studio-recorded speech (even translated ones) is later matched to the acoustic environment shown on screen. In games and audio-based virtual reality, AEM enables sounds to match the visual or imagined environment experienced by the user. At the same time, as AEM techniques become increasingly powerful, arbitrary ``relocation” of audio introduces serious concerns for service providers and content owners. With modern audio generative models, it is possible to synthesize a person’s voice and insert it into an existing recording, e.g., a meeting or an oral presentation. AEM further amplifies this risk by making such insertions sound as if they were recorded in the same physical space, misleading listeners to trust the modified audio. Conversely, benign users may also wish to embed information into room effects, such as physical location, recording time, event metadata, or copyright ownership. This naturally connects AEM with audio watermarking \cite{wm-audioseal,wm-timbre,wm-wavmark}, which aims to embed information into audio signals without perceptually degrading quality. 

\begin{figure}[t]
    \centering
    \includegraphics[width=\linewidth]{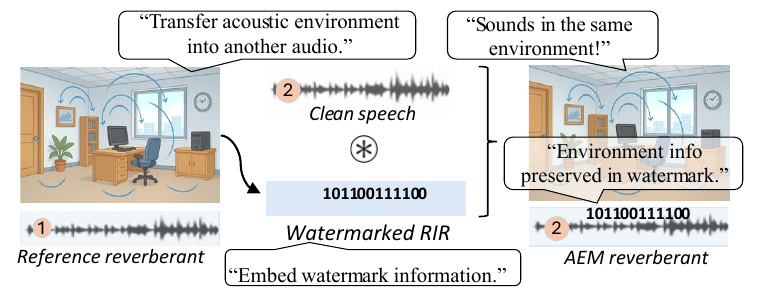}
    \caption{EchoMark applies acoustic environments from reference reverberant to clean speech, creating the same room effect. Meanwhile, a watermark is embedded in the RIR, whose bits can be detected in the transferred audio.}
    \label{fig:intro}
    \vspace{-4mm}
\end{figure}

Despite this clear need, no existing system jointly supports high-quality AEM and reliable watermarking. This motivates us to propose the first AEM watermarking framework, \textbf{EchoMark} (illustrated in Fig.~\ref{fig:intro}), allowing users to enjoy creative environment transfer while maintaining responsibility and traceability. 
Prior studies \cite{aem-perceptual,aem-perceptual-mit,model-fins} show that the primary component governing perceived room characteristics is the room impulse response (RIR), whose waveform represents the time-of-arrivals and strengths of sound reflections from the source. In other words, the quality of AEM can be interpreted as \textbf{how closely a reconstructed RIR matches the ground truth}. 
In parallel, as for watermarking, the key problem lies in \textbf{how to identify degrees of freedom that are imperceptible to the human auditory system (HAS)}. As an answer to joint AEM and watermarking, we propose to leverage the representation power of deep neural networks to reconstruct and embed imperceptible information into variant-length and decaying RIR waveforms.

\smallskip
\noindent{\textbf{Challenges.}}
Overall, we face three major challenges and address them accordingly.
(1) \textbf{Architecture Design.} If reconstruction and watermarking are handled separately, it becomes non-ideal for models to accommodate variable-length RIRs in the middle. In addition, we prefer EchoMark to directly generate only watermarked RIRs (wRIRs), avoiding redundancy that first estimate a non-watermarked RIR and then generate another watermarked version.
(2) \textbf{Capacity Exploitation.} RIR waveforms generators are unlike those for audio, such as HiFiGAN~\cite{model-hifigan}. Instead, prior works~\cite{model-fins,model-decor} show that physics motivated inductive biases can greatly improve generation efficiency while maintaining perceptual quality. However, fewer learnable parameters also imply a smaller degree of freedom for watermark embedding, creating a tension between RIR generation fidelity and watermark capacity.
(3) \textbf{Post-Convolution Detection.} Unlike conventional audio watermarking (e.g., \cite{wm-audioseal}), where embedding and detection are performed directly on the waveform, our RIR watermark is first convolved with arbitrary clean speech to simulate room effects and is then detected from the resulting reverberant speech. This requires the watermark to be embedded in robust features rather than fragile signal details in order to survive the AEM process.

To overcome these challenges, we propose an \textbf{EWG-D} architecture. The \textbf{\underline{E}}ncoder consumes reference reverberant speech and produces a fixed-length latent embedding. The \textbf{\underline{W}}atermarker fuses message embeddings into the RIR embedding generated by \textbf{\underline{E}}. The \textbf{\underline{G}}enerator then outputs the wRIR waveform, which is later convolved with arbitrary clean speech during AEM process. Next, based on human perception of RIRs, we incorporate perceptually-insignificant inductive biases into \textbf{\underline{G}} to increase effective watermark capacity. Finally, to ensure that watermarking operates on human sense-related factors, the post-convolution \textbf{\underline{D}}etector is trained simulating the exact AEM process.

\smallskip
\noindent{\textbf{Use Cases}.}
EchoMark enables several practical scenarios in the era of immersive auditory virtual reality:
(1) \underline{\textit{Remote participation.}} When recording online meetings, remote speakers can be matched to the same room acoustics as on-site participants for better experiences, with a mandatory watermark embedded within AEM-transferred speech.
(2) \underline{\textit{Environment-aware effects.}} Natural RIRs can be replaced with wRIRs marked by physical location, which could be retrieved afterwards.
(3) \underline{\textit{Understandable blind watermarking.}} EchoMark can even be viewed as blind audio watermarking, where the intentionally designed room effects announce the presence of watermark.

\section{Background}
\noindent{\textbf{Room Acoustics.}}
An RIR is commonly decomposed into a direct path, sparse early reflections, and dense late reverberation with exponential decay. Parameters derived from these components, such as reverberation time (RT) and direct-to-reverberant ratio (DRR), effectively characterize perceived room properties \cite{aem-perceptual}. Recent studies further identify the delay–frequency profile as perceptually salient, while showing limited listener sensitivity to higher-order decay rates and moderate frequency-dependent variations for complex signals like speech \cite{aem-perceptual-mit}.

\smallskip
\noindent{\textbf{RIR Estimation.}}
Recent AEM methods mainly follow two paradigms: estimating RIRs directly from reverberant speech and generating audio conditioned on learned RIR representations. Early signal processing techniques, such as WPE \cite{method-wpe} and NMF \cite{method-NMF}, have largely been replaced by neural approaches that jointly model dereverberation and RIR estimation \cite{model-buddy}.
More recent work extracts compact RIR embeddings to condition speech generators for environment transfer \cite{riremb-su,riremb-koo}. Generative models such as IR-GAN \cite{rirest-irgan} and REC-RIR \cite{model-rec-rir} synthesize plausible RIRs from reverberant input, while FiNS \cite{model-fins} and DECOR \cite{model-decor} introduce physical inductive biases to model late reverberation via noise filtering. Informed RIR estimation further incorporates visual cues, highlighting the advantages of multi-modal context when available.

\smallskip
\noindent{\textbf{Audio Watermarking.}}
Deep learning has advanced audio watermarking by enabling robust detection and message embedding with minimal perceptual degradation. Existing methods support applications such as copyright protection and content authentication, using frequency-domain modeling \cite{wm-dear,wm-timbre}, tamper localization \cite{wm-audioseal}, or invertible networks \cite{wm-wavmark}. However, these approaches focus on speech content or speaker identity and do not address watermarking room impulse responses. Since the RIR encodes the identity of the acoustic space in AEM, watermarking the RIR offers a natural and proactive solution to security and accountability concerns in environment-transferred audio.

\section{Problem Formulation}
\noindent{\textbf{Natural Reverberant Signal.}}
Natural reverberant speech can be modeled as the convolution of clean speech with a room impulse response, together with additive noise. This process can be represented in both the time domain and the time--frequency domain using the convolutive transfer function (CTF) approximation:
\begin{equation}
\label{eq:RIR-covolution}
    Y(t,f) = \sum_{\tau=0}^{T-1} X(t - \tau, f) H(\tau, f) + W(t,f),
\end{equation}
where \(X\), \(Y\), \(H\), and \(W\) denote the time--frequency representations of clean speech, reverberant speech, RIR, and additive noise at frame \(t\) and frequency bin \(f\). Convolution with the RIR causes the energy of clean speech to spread across successive frames\footnote{We omit time and frequency indices for simplicity later in this paper.}. Therefore, we adopt log-scale spectrograms as model input, which better represent the frequency-dependent exponential decay characteristics of RIRs.

\smallskip
\noindent{\textbf{Watermarked RIR Generation.}}
The task of EchoMark is to generate a watermarked RIR \(\hat{h}\) that is perceptually similar to the original \(h\), where \(h\) denotes the time-domain representation of \(H\). First, an encoder consumes a reference reverberant waveform and produces a fixed-length latent embedding. This embedding serves as a compact representation of the acoustic environment, avoiding explicit estimation of variable-length RIR waveforms. Next, the message embedding \(M(m)\) is merged with the RIR embedding to generate the watermarked RIR \(\hat{h}\). This process is constrained by both perceptual similarity and watermark detection accuracy.
\begin{equation}
\label{eq:goal-generator}
    \hat{h} = G(E(Y), M(m)) \quad \text{s.t.} \quad P(\hat{h}) \approx P(h),~D(Y') = m,
\end{equation}
where \(E\) denotes the RIR encoder, \(W\) the watermarker, and \(G\) the RIR generator. \(P(\cdot)\) represents a perceptual evaluation of the acoustic environment, and \(D(\cdot)\) is the watermark detector applied to the generated reverberant signal \(Y' = \mathrm{STFT}(x' * \hat{h})\), with \(*\) denoting convolution.

\smallskip
\noindent{\textbf{Watermark Detection on AEM Speech.}}
The goal of this step is to verify whether a given reverberant speech signal \(Y_\text{test}\) was generated by the service provider's AEM model, i.e., EchoMark. The watermark detector \(D(\cdot)\) is applied to the environment-transferred input and produces:
\begin{equation}
\label{eq:goal-detector}
D(\tilde{Y}) =
\begin{cases}
    (1, m), & \text{if } Y_\text{test} \in \hat{Y}, \\
    (0, z), & \text{otherwise},
\end{cases}
\end{equation}
where the output \((1, m)\) indicates the presence of a valid watermark along with the decoded message \(m\) of \(M\) bits. If the watermark is not detected, the detector returns \((0, z)\), where \(z\) is an arbitrary output to be discarded.

\begin{figure}[t]
    \centering
    \includegraphics[width=.9\linewidth]{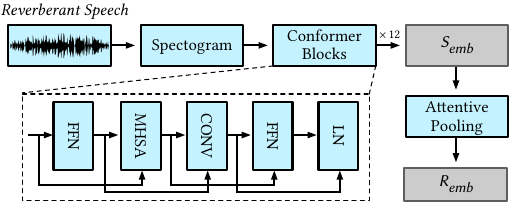}
    \caption{\textbf{EchoMark RIR Encoder.} FFN: Feed-Forward Network; MHSA: Multi-Head Self-Attention; CONV: Convolutional Layer; LN: LayerNorm.}
    \label{fig:encoder}
    \vspace{-4mm}
\end{figure}

\section{EchoMark Design}
\label{sec:model}

\noindent{\textbf{RIR Encoder.}}
We adopt the Conformer architecture~\cite{model-conformer} as the backbone of the RIR encoder, as it is effective at modeling both local spectral patterns and long-range temporal dependencies in audio. The overall architecture is illustrated in Fig.~\ref{fig:encoder}. To aggregate frame-level representations into a fixed-length embedding, we apply attentive pooling, following the design used in ECAPA-TDNN~\cite{model-ecapa}.

The encoding process is defined as
\begin{equation}
\begin{aligned}
    S_{\text{emb}} &= \mathrm{Conformer}(Y),\\
    R_{\text{emb}} &= \mathrm{AttnPool}(S_{\text{emb}}),
\end{aligned}
\end{equation}
where \(Y \in \mathbb{R}^{T \times F}\) denotes the log-magnitude spectrogram of the input reverberant speech, \(S_{\text{emb}} \in \mathbb{R}^{T \times d}\) is the sequence of latent representations, and \(R_{\text{emb}} \in \mathbb{R}^{d}\) is the resulting fixed-length RIR embedding that summarizes the acoustic environment.

\begin{figure}[t]
    \centering
    \includegraphics[width=1\linewidth]{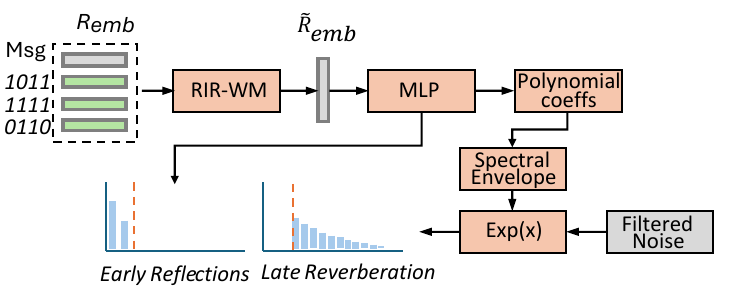}
    \caption{\textbf{EchoMark RIR Watermarker and Generator.} Early: MLP generated from watermarked embedding; Late: learned polynomial decay envelope in log-scale for noise shaping.}
    \label{fig:generator}
\end{figure}

\begin{figure}[t]
    \centering
    \includegraphics[width=.9\linewidth]{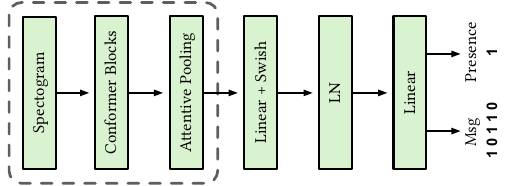}
    \caption{\textbf{EchoMark Watermark Detector.} Encoder architecture reused. Logits output for presence and message detection, respectively.}
    \label{fig:detector}
    \vspace{-4mm}
\end{figure}

\smallskip
\noindent{\textbf{RIR Watermarker.}} As illustrated in Fig.~\ref{fig:generator}, EchoMark embeds watermark information directly into the RIR embedding space. Given an \(M\)-bit binary message \(m \in \mathbb{B}^{M}\), we sequentially partition it into \(q\)-bit chips \(\{c_1, \ldots, c_n\}\), where \(n = M/q=3\), when \(M=12, q=4\)). Each chip is independently projected into the latent space through a fully connected layer. These projections are then concatenated with the original RIR embedding and fused as
\begin{equation}
    \tilde{R}_{\text{emb}} =
    \mathrm{FC}\big(
    \mathrm{Concat}(R_{\text{emb}}, \mathrm{FC}(c_1), \ldots, \mathrm{FC}(c_n))
    \big).
\end{equation}
The resulting \(\tilde{R}_{\text{emb}}\) serves as the watermarked acoustic representation used for subsequent RIR generation. Unless otherwise stated, we report results with a watermark capacity of \(M=12\) and chip size \(q=4\).

\smallskip
\noindent{\textbf{RIR Generator.}}
Room impulse responses exhibit markedly different behaviors between early reflections and late reverberation: early reflections are sparse and impulsive, while late reverberation is dense and characterized by approximately exponential energy decay. Motivated by classical room acoustics theory and recent neural RIR models~\cite{theory-rir-model,model-fins,model-decor}, we explicitly model these two components separately.
In this work, we focus on perceptual similarity under mono-channel listening conditions. Although early reflections encode spatial cues in binaural settings, their perceptual impact is relatively limited in the mono case\footnote{As shown in~\cite{aem-perceptual-mit}, replacing early reflections leads to negligible perceptual differences for complex sources, including speech signals.}. Accordingly, we generate the first 50~ms of early reflections directly from \(\tilde{R}_{\text{emb}}\) using MLP.

For late reverberation, we follow the successful paradigm of noise-based synthesis with explicit exponential decay used in FiNS~\cite{model-fins} and DECOR~\cite{model-decor}, while introducing modifications to improve both AEM quality and watermark capacity. Specifically, we apply linear bandpass filters with a 20~Hz guard interval as learnable noise-shaping filters to model frequency-dependent reverberation behavior. Unlike prior approaches that employ shared decay bases across bands, we parameterize each band with a fixed-order polynomial decay, where higher-order coefficients (\(n > 1\)) are constrained to small magnitudes (max set to 0.01). This design provides additional degrees of freedom that are largely imperceptible to human listeners, making them suitable for robust watermark embedding.

The generation process is formulated as
\begin{equation}
\begin{aligned}
    \hat{h}_{\text{early}} &= \mathrm{MLP}(\tilde{R}_{\text{emb}}),\\
    \hat{h}_{\text{late}} &= \sum_{f} \exp\!\left(b_0 + \sum_{n} a_n t^n \right) w_f(t),
\end{aligned}
\end{equation}
where \(\hat{h}_{\text{early}}\) denotes the early reflection component within the first 50~ms, and \(\hat{h}_{\text{late}}\) represents the late reverberation that follows. To ensure a physically plausible energy relationship between the separately generated early and late components, we explicitly fix the first sample of the complete RIR \(\hat{h}\) to `1' and constrain all remaining samples in both branches to have magnitudes smaller than one. This normalization enforces a natural direct-path dominance and stabilizes the relative energy decay across the RIR.

\smallskip
\noindent{\textbf{Watermark Detector.}}
To verify the embedded watermark, we employ a detector that shares the same network architecture as the RIR Encoder but is trained with independent parameters, as shown in Fig.~\ref{fig:detector}. After attentive pooling, the latent representation is passed through a linear layer with Swish activation, followed by a final linear projection. The detector outputs a vector of dimension \(M+1\), where the first element indicates watermark presence and the remaining \(M\) elements correspond to the recovered message bits. During training, raw outputs are used directly, while inference applies zero-thresholding to obtain binary decisions.

\section{Model Training}
\label{sec:training}

\subsection{Data Preparation and Augmentation}
Training data are constructed by convolving clean speech with real RIRs. Therefore, ground-truth RIR can provide direct supervision with reconstruction losses. Both clean speech and RIR waveforms are clipped to 2 seconds, and the resulting reverberant audio is truncated to capture long-range reverberation patterns.
To improve robustness, input reverberant speech is augmented with isotropic noise at randomly sampled SNRs between 5 and 20 dB. During training, the generated watermarked RIR is convolved with randomly shuffled clean speech to train the watermark detector under varying speech content. Reverberant speech generated using real RIRs is also included for balanced watermark presence detection.

\subsection{Loss Functions}
EchoMark is trained to jointly optimize AEM quality and watermark detection. The total loss is defined as
\begin{equation}
    \mathcal{L}_{\text{total}} = \mathcal{L}_{\text{RIR}} + \mathcal{L}_{\text{WM}}.
\end{equation}

\noindent{\textbf{RIR Perceptual Loss.}}
We adopt a combination of multi-scale spectral (MSS) loss and energy decay convergence (EDC) loss \cite{rir-loss} to encourage perceptual similarity between generated and target RIRs:
\begin{equation}
\label{eq:rir-loss}
    \mathcal{L}_{\text{RIR}} = \mathcal{L}_{\text{MSS}} + \mathcal{L}_{\text{EDC}}.
\end{equation}

\noindent{\textbf{Watermarking Loss.}}
Watermark presence detection and message decoding are supervised using a margin-based hinge loss over \(M+1\) outputs, where the first element corresponds to watermark presence and the remaining \(M\) elements represent message bits:
\begin{equation}
\label{eq:wm-loss}
    \mathcal{L}_{\text{WM}} =
    \frac{1}{M+1} \sum_{i=0}^{M}
    \max\left(0, 1 - y_i \cdot \hat{y}_i\right),
\end{equation}
where \(y_i \in \{-1,1\}\) denotes the ground-truth label and \(\hat{y}_i\) is the corresponding model output.

\smallskip
\noindent{\textbf{Optimization.}}
All components are trained jointly using the AdamW optimizer \cite{method-adamW} with a learning rate of \(1\times10^{-4}\) and batch size 16.

\begin{figure}[t]
    \centering
    \includegraphics[width=.9\linewidth]{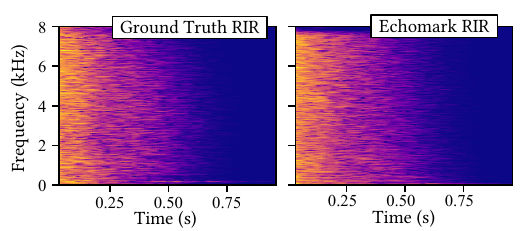}
    \caption{Ground truth (left) and EchoMark RIR (right) spectrograms.}
    \vspace{-4mm}
    \label{fig:rir-stft}
\end{figure}

\section{Experiment Setup}

\subsection{Dataset}
We use LibriSpeech \cite{data-librispeech} as the source of clean speech. Target room impulse responses are constructed by merging real-world RIR datasets from RWCP \cite{data-rwcp}, the REVERB Challenge \cite{data-rvb}, AIR \cite{data-air}, and BUT Reverb \cite{data-BUT}. For each dataset, RIRs are split into training and testing sets with an 80:20 ratio. Corresponding subsets of LibriSpeech are used as clean speech for training and evaluation. All clean and reverberant speech segments are truncated or padded to 2 seconds during training.

\subsection{Metrics}
\noindent{\textbf{Room Acoustics.}}
We evaluate acoustic similarity using reverberation time \(RT_{60}\), defined as the time required for the RIR energy to decay by 60~dB; the direct-to-reverberant ratio (DRR), which measures the energy ratio between direct sound and reverberation; and clarity (\(C_{50}\)), which quantifies the energy ratio between early and late reflections. For all metrics, we report the root mean squared error (RMSE) with respect to the ground truth and compare against other RIR generation models. In addition, we conduct a subjective evaluation using mean opinion score (MOS), where listeners are presented with ground-truth and AEM-generated audio and asked to rate their similarity in terms of acoustic environment on a scale from 0 (dissimilar) to 5 (identical).

\smallskip
\noindent{\textbf{Watermarking.}}
We evaluate watermarking performance using detection (presence) accuracy  and message decoding accuracy. Watermark presence detection is measured over 168 batches, with each sample having a 50\% probability of being watermarked, resulting in a balanced (50:50) evaluation setting. Message decoding accuracy is evaluated only on samples that contain valid watermarks. Overall, decoding accuracy is computed over a total of
\(12~\text{(bits/RIR)} \times 16~\text{(batch size)} \times 168~\text{(batches)} = 32{,}256\)
embedded bits.

\begin{figure}[t]
    \centering
    \includegraphics[width=0.85\linewidth]{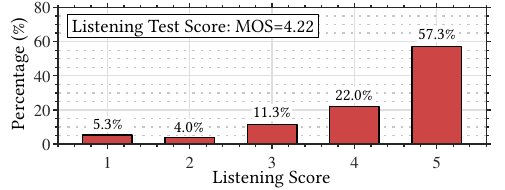}
    \caption{Listening test and mean opinion score.}
    \label{fig:MOS}
    \vspace{-4mm}
\end{figure}

\vspace{-2mm}
\section{Results}
We first visualize an example of generated watermarked RIR in Fig.~\ref{fig:rir-stft} and compare it with its real-world counterpart. Benefiting from similar physics-inspired inductive biases~\cite{model-fins}, the generated RIRs exhibit energy decay patterns over time that closely match those of real RIRs across most frequency bands. In addition, the use of spectral-domain losses leads to similar time-frequency structures, resulting in perceptually consistent reverberation effects when applied to test audio.

\vspace{-2mm}
\subsection{AEM Quality}
\noindent{\textbf{Objective Evaluation.}}
As shown in Table~\ref{tab:rir_metrics}, EchoMark demonstrates competitive performance in blind RIR reconstruction from reverberant audio when compared with state-of-the-art methods~\cite{model-fins,model-buddy,model-rec-rir}. For \(RT_{60}\), EchoMark exhibits a slightly higher RMSE than FiNS, with only a small performance gap. For DRR and \(C_{50}\), EchoMark also achieves satisfactory results, trailing only the strongest baseline (Rec-RIR). We emphasize that EchoMark is not designed to outperform dedicated RIR reconstruction methods. Instead, the primary goal is to embed watermarks without degrading perceptual AEM quality. The results confirm that watermark embedding does not introduce noticeable degradation in room acoustic characteristics\footnote{This work focuses on mono-channel perceptual similarity and does not explicitly model spatial cues in early reflections for multi-channel audio.}.

\begin{table}[h]
\centering
\vspace{-4mm}
\caption{RMSE of Room Acoustics across RIR Generation Methods.}
\label{tab:rir_metrics}
\begin{tabular}{lcccc}
\toprule
\textbf{Methods} & \textbf{FiNS\cite{model-fins}} & \textbf{BUDDy\cite{model-buddy}} & \textbf{Rec-RIR\cite{model-rec-rir}} & \textbf{EchoMark} \\
\midrule
$\epsilon_\text{RT60}$ (s)   & 0.167 & 0.166 & 0.104 & 0.188 \\
$\epsilon_\text{DRR}$ (dB)        & 2.639  & 4.360  & 0.794  & 2.590  \\
$\epsilon_\text{C50}$ (dB)        & 6.597  & 4.477  & 1.019  & 1.503  \\
\bottomrule
\vspace{-4mm}
\end{tabular}
\end{table}

\begin{figure}[t]
  \centering
  \begin{minipage}[b]{0.49\linewidth}
    \centering
    \includegraphics[width=\linewidth]{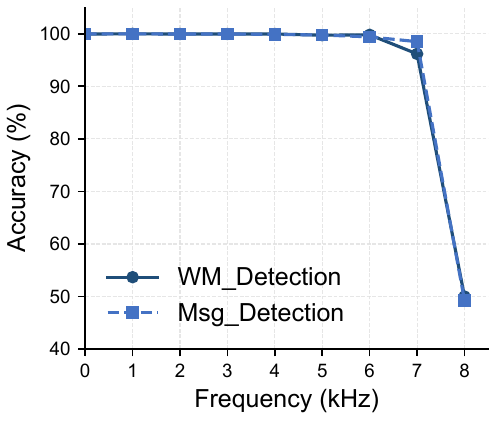}
  \end{minipage}
  \begin{minipage}[b]{0.49\linewidth}
    \centering
    \includegraphics[width=\linewidth]{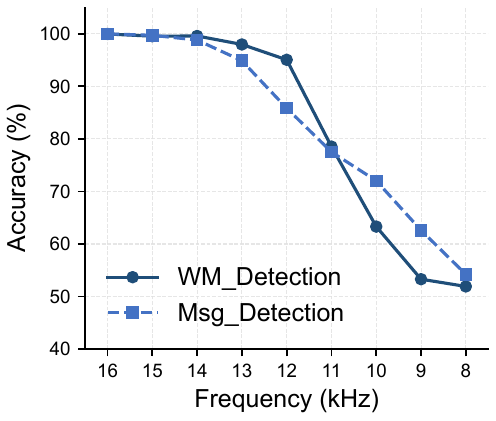}
  \end{minipage}
  
  \caption{Overall detection performance across various frequencies (Left: High-Pass Filter, Right: Low-Pass Filter).} \vspace{-4mm}
  \label{fig:WM_vs_freq}
\end{figure}

\smallskip
\noindent{\textbf{Subjective Evaluation.}}
We conduct a human listening test on 15 EchoMark-generated reverberant utterances, each evaluated against its ground-truth counterpart by 20 participants. As shown in Fig.~\ref{fig:MOS}, EchoMark achieves a mean opinion score (MOS) of 4.22. Moreover, 57.3\% of samples receive the highest rating, and 79.3\% are rated above the neutral midpoint (score \(>3\)), indicating strong perceptual alignment in acoustic environment matching.

\subsection{Watermark Performance.}
\noindent{\textbf{Capacity.}}
EchoMark embeds 12 bits per RIR, which remain detectable after convolution with speech. For reference, AudioSeal~\cite{wm-audioseal} and WavMark~\cite{wm-wavmark} embed 16 bits directly into speech utterances, which usually have longer duration.

\noindent{\textbf{Accuracy \& Robustness.}}
We report watermark accuracy and robustness in Table~\ref{tab:wm_robustness}. EchoMark maintains reliable decoding when additive noise is applied to both reference and AEM-transferred reverberant signals, with no noticeable degradation as SNR decreases from 20~dB to 5~dB. We further evaluate several attacks designed to remove the watermark without perceptually altering room effects. Among all tested attacks, only resampling to 8~kHz fails due to anti-aliasing filtering, while EchoMark remains robust under other attacks.

\begin{table}[h]
\centering
\vspace{-3mm}
\caption{Watermark robustness under noise and signal perturbations.}
\label{tab:wm_robustness}
\begin{tabular}{lccccc}
\toprule
\textbf{SNR (dB)} & \textbf{5} & \textbf{10} & \textbf{15} & \textbf{20} & \textbf{$\infty$} \\
\midrule
Presence (\%) & 99.61 & 99.65 & 99.85 & 99.96 & 99.96 \\
Message (\%)  & 98.12  & 99.15  & 99.65  & 99.81  & 99.99  \\
\midrule
\textbf{Attacks} & \textbf{SF} & \textbf{RA} & \textbf{DC} & \textbf{RS} & \textbf{DB} \\
\midrule
Presence (\%) & 99.96 & 100 & 99.99 & 51.92 & 99.96 \\
Message (\%)  & 99.97  & 99.98  & 99.98  & 54.22  & 99.51  \\
\bottomrule
\end{tabular}

\vspace{1mm}
\begin{flushleft} 
{\scriptsize
SF: Sign Flip; RA: Random Amplitude Scale [0.25-1.75]; DC: Drop Chunk [0.25-0.5]; RS: Resample [16kHz-8kHz-16kHz]; DB: Drop Bit Resolution [float32-int8-float32].}
\end{flushleft}
\vspace{-4mm}
\end{table}

\noindent{\textbf{Watermark on Frequencies.}}
We further analyze the frequency components that carry the embedded watermark by applying bandpass filters to the watermarked RIRs and evaluating decoding performance. As shown in Fig.~\ref{fig:WM_vs_freq}, removing low-frequency components does not affect watermark recovery until approximately 7~kHz. Conversely, applying low-pass filtering begins to degrade performance above 13~kHz. These results indicate that EchoMark primarily embeds watermark information within the 7–13~kHz frequency range.

\begin{table}[h]
\centering
\vspace{-3mm}
\caption{Watermark robustness under different test conditions.}
\label{tab:wm_generalization}
\begin{tabular}{lccc}
\toprule
\textbf{Condition} & \textbf{Unseen RIR} & \textbf{Language} & \textbf{Music} \\
\midrule
$\epsilon_{RT60}$ & 0.195 & 0.189 & 0.203 \\
Message Acc (\%)     & 99.97 & 99.96 & 99.53 \\
\bottomrule
\end{tabular}
\vspace{-3mm}
\end{table}

\subsection{Generalization.}
Finally, we test how the model can generalize to unseen RIR (synthesized), different language, and working on music. Results in Table~\ref{tab:wm_generalization} show satisfactory results in room acoustics and watermark performance.

\section{Conclusion}
We present EchoMark, the first deep learning framework that generates watermarked RIR from reverberant speech for AEM. EchoMark integrates an RIR encoder, watermarker, generator, and detector in a unified architecture jointly optimized for RIR reconstruction and watermark embedding. By leveraging higher-order polynomial coefficients as decay envelopes, EchoMark achieves a 12-bit watermark capacity without noticeably degrading room acoustic effects, enabling responsible AEM experiences for virtual reality and perceptually interpretable blind watermarking.
Experiments demonstrate that EchoMark achieves room acoustic accuracy comparable to strong RIR reconstruction baselines that do not support watermarking, as confirmed by human listening tests. Moreover, EchoMark attains over 99\% accuracy in both watermark detection and message decoding, and remains robust under noise and watermark removal attempts, highlighting a practical path toward secure and perceptually faithful acoustic environment transfer in multimedia systems.

\bibliographystyle{IEEEbib}
\bibliography{icme2026references}

@article{rirest-irgan,
  title={IR-GAN: Room impulse response generator for far-field speech recognition},
  author={Ratnarajah, Anton and Tang, Zhenyu and Manocha, Dinesh},
  journal={arXiv preprint arXiv:2010.13219},
  year={2020}
}

@inproceedings{riremb-koo,
  title={Reverb conversion of mixed vocal tracks using an end-to-end convolutional deep neural network},
  author={Koo, Junghyun and Paik, Seungryeol and Lee, Kyogu},
  booktitle={ICASSP 2021-2021 IEEE international conference on acoustics, speech and signal processing (ICASSP)},
  pages={81--85},
  year={2021},
  organization={IEEE}
}

@inproceedings{riremb-su,
  title={Acoustic matching by embedding impulse responses},
  author={Su, Jiaqi and Jin, Zeyu and Finkelstein, Adam},
  booktitle={ICASSP 2020-2020 IEEE international conference on acoustics, speech and signal processing (ICASSP)},
  pages={426--430},
  year={2020},
  organization={IEEE}
}

@article{aem-perceptual,
  title={Perceptual matching of room acoustics for auditory augmented reality in small rooms-literature review and theoretical framework},
  author={Neidhardt, Annika and Schneiderwind, Christian and Klein, Florian},
  journal={Trends in Hearing},
  volume={26},
  pages={23312165221092919},
  year={2022},
  publisher={SAGE Publications Sage CA: Los Angeles, CA}
}

@article{aem-perceptual-mit,
  title={Statistics of natural reverberation enable perceptual separation of sound and space},
  author={Traer, James and McDermott, Josh H},
  journal={Proceedings of the National Academy of Sciences},
  volume={113},
  number={48},
  pages={E7856--E7865},
  year={2016},
  publisher={National Academy of Sciences}
}

@inproceedings{wm-audioseal,
  title={Proactive Detection of Voice Cloning with Localized Watermarking},
  author={San Roman, Robin and Fernandez, Pierre and Elsahar, Hady and D{\'e}fossez, Alexandre and Furon, Teddy and Tran, Tuan},
  booktitle={International Conference on Machine Learning},
  pages={43180--43196},
  year={2024},
  organization={PMLR}
}

@inproceedings{wm-timbre,
  title={Detecting Voice Cloning Attacks via Timbre Watermarking},
  author={Liu, Chang and Zhang, Jie and Zhang, Tianwei and Yang, Xi and Zhang, Weiming and Yu, Nenghai},
  booktitle={NDSS},
  year={2024}
}

@article{wm-wavmark,
  title={Wavmark: Watermarking for audio generation},
  author={Chen, Guangyu and Wu, Yu and Liu, Shujie and Liu, Tao and Du, Xiaoyong and Wei, Furu},
  journal={arXiv preprint arXiv:2308.12770},
  year={2023}
}

@inproceedings{wm-dear,
  title={Dear: A deep-learning-based audio re-recording resilient watermarking},
  author={Liu, Chang and Zhang, Jie and Fang, Han and Ma, Zehua and Zhang, Weiming and Yu, Nenghai},
  booktitle={Proceedings of the AAAI Conference on Artificial Intelligence},
  volume={37},
  number={11},
  pages={13201--13209},
  year={2023}
}

@article{model-conformer,
  title={Conformer: Convolution-augmented transformer for speech recognition},
  author={Gulati, Anmol and Qin, James and Chiu, Chung-Cheng and Parmar, Niki and Zhang, Yu and Yu, Jiahui and Han, Wei and Wang, Shibo and Zhang, Zhengdong and Wu, Yonghui and others},
  journal={arXiv preprint arXiv:2005.08100},
  year={2020}
}

@article{model-ecapa,
  title={Ecapa-tdnn: Emphasized channel attention, propagation and aggregation in tdnn based speaker verification},
  author={Desplanques, Brecht and Thienpondt, Jenthe and Demuynck, Kris},
  journal={arXiv preprint arXiv:2005.07143},
  year={2020}
}

@inproceedings{model-fins,
  title={Filtered noise shaping for time domain room impulse response estimation from reverberant speech},
  author={Steinmetz, Christian J and Ithapu, Vamsi Krishna and Calamia, Paul},
  booktitle={2021 IEEE workshop on applications of signal processing to audio and acoustics (WASPAA)},
  pages={221--225},
  year={2021},
  organization={IEEE}
}

@article{model-hifigan,
  title={Hifi-gan: Generative adversarial networks for efficient and high fidelity speech synthesis},
  author={Kong, Jungil and Kim, Jaehyeon and Bae, Jaekyoung},
  journal={Advances in neural information processing systems},
  volume={33},
  pages={17022--17033},
  year={2020}
}

@article{model-decor,
  title={Deep room impulse response completion},
  author={Lin, Jackie and G{\"o}tz, Georg and Schlecht, Sebastian J},
  journal={EURASIP Journal on Audio, Speech, and Music Processing},
  volume={2025},
  number={1},
  pages={20},
  year={2025},
  publisher={Springer}
}

@article{theory-rir-model,
  title={Fifty years of artificial reverberation},
  author={Valimaki, Vesa and Parker, Julian D and Savioja, Lauri and Smith, Julius O and Abel, Jonathan S},
  journal={IEEE Transactions on Audio, Speech, and Language Processing},
  volume={20},
  number={5},
  pages={1421--1448},
  year={2012},
  publisher={IEEE}
}

@ARTICLE{method-wpe,
  author={Nakatani, Tomohiro and Yoshioka, Takuya and Kinoshita, Keisuke and Miyoshi, Masato and Juang, Biing-Hwang},
  journal={IEEE Transactions on Audio, Speech, and Language Processing}, 
  title={Speech Dereverberation Based on Variance-Normalized Delayed Linear Prediction}, 
  year={2010},
  volume={18},
  number={7},
  pages={1717-1731},
  doi={10.1109/TASL.2010.2052251}}

@inproceedings{method-NMF,
  title={Joint acoustic and spectral modeling for speech dereverberation using non-negative representations},
  author={Mohammadiha, Nasser and Smaragdis, Paris and Doclo, Simon},
  booktitle={2015 IEEE international conference on acoustics, speech and signal processing (ICASSP)},
  pages={4410--4414},
  year={2015},
  organization={IEEE}
}

@inproceedings{data-librispeech,
  title={Librispeech: an asr corpus based on public domain audio books},
  author={Panayotov, Vassil and Chen, Guoguo and Povey, Daniel and Khudanpur, Sanjeev},
  booktitle={2015 IEEE international conference on acoustics, speech and signal processing (ICASSP)},
  pages={5206--5210},
  year={2015},
  organization={IEEE}
}

@inproceedings{data-rwcp,
  title={Acoustical Sound Database in Real Environments for Sound Scene Understanding and Hands-Free Speech Recognition.},
  author={Nakamura, Satoshi and Hiyane, Kazuo and Asano, Futoshi and Nishiura, Takanobu and Yamada, Takeshi},
  booktitle={LREC},
  year={2000}
}

@incollection{data-rvb,
  title={The REVERB challenge: A benchmark task for reverberation-robust ASR techniques},
  author={Kinoshita, Keisuke and Delcroix, Marc and Gannot, Sharon and Habets, Emanu{\"e}l AP and Haeb-Umbach, Reinhold and Kellermann, Walter and Leutnant, Volker and Maas, Roland and Nakatani, Tomohiro and Raj, Bhiksha and others},
  booktitle={New Era for Robust Speech Recognition: Exploiting Deep Learning},
  pages={345--354},
  year={2017},
  publisher={Springer}
}

@inproceedings{data-air,
  title={A binaural room impulse response database for the evaluation of dereverberation algorithms},
  author={Jeub, Marco and Schafer, Magnus and Vary, Peter},
  booktitle={2009 16th international conference on digital signal processing},
  pages={1--5},
  year={2009},
  organization={IEEE}
}

@article{data-BUT,
  title={Building and evaluation of a real room impulse response dataset},
  author={Sz{\"o}ke, Igor and Sk{\'a}cel, Miroslav and Mo{\v{s}}ner, Ladislav and Paliesek, Jakub and {\v{C}}ernock{\`y}, Jan},
  journal={IEEE Journal of Selected Topics in Signal Processing},
  volume={13},
  number={4},
  pages={863--876},
  year={2019},
  publisher={IEEE}
}

@article{method-adamW,
  title={Decoupled weight decay regularization},
  author={Loshchilov, Ilya and Hutter, Frank},
  journal={arXiv preprint arXiv:1711.05101},
  year={2017}
}

@inproceedings{rir-loss,
author = {Dal Santo, Gloria and Prawda, Karolina and Schlecht, Sebastian and Välimäki, Vesa},
year = {2024},
month = {10},
pages = {1409-1413},
title = {Similarity Metrics for Late Reverberation},
doi = {10.1109/IEEECONF60004.2024.10943013}
}

@article{model-buddy,
  title={Unsupervised blind joint dereverberation and room acoustics estimation with diffusion models},
  author={Lemercier, Jean-Marie and Moliner, Eloi and Welker, Simon and V{\"a}lim{\"a}ki, Vesa and Gerkmann, Timo},
  journal={IEEE Transactions on Audio, Speech and Language Processing},
  year={2025},
  publisher={IEEE}
}

@article{model-rec-rir,
  title={Rec-RIR: Monaural Blind Room Impulse Response Identification via DNN-based Reverberant Speech Reconstruction in STFT Domain},
  author={Wang, Pengyu and Li, Xiaofei},
  journal={arXiv preprint arXiv:2509.15628},
  year={2025}
}

\end{document}